\documentclass[aps,preprint,showpacs,showkeys]{revtex4}
\usepackage{amsmath,amssymb}
\usepackage{bm}
\usepackage[dvips]{graphicx}

\begin{document}

\title{Bulk and single-particle properties of hyperonic matter at finite
temperature}

\author{A. Rios$^1$, A. Polls$^1$, A. Ramos$^1$ and I. Vida\~na$^2$}

\affiliation{$^1$Departament d'Estructura i Constituents de la
Mat\`eria, Universitat de Barcelona, Avda. Diagonal 647, E-08028
Barcelona, Spain}

\affiliation{$^2$Gesellschaft f\"{u}r Schwerionenforschung (GSI),
Planckstrasse. 1, D-64291 Darmstadt, Germany}

\begin{abstract}

Bulk and single-particle properties of hot hyperonic matter are studied
within the Brueckner--Hartree--Fock approximation extended to finite
temperature.
The bare interaction in the nucleon sector is the Argonne V18
potential supplemented with an effective three-body force to
reproduce the saturating properties of nuclear matter. The modern
Nijmegen NSC97e potential is employed for the hyperon-nucleon and
hyperon-hyperon interactions. The effect of the temperature on the
in-medium effective interaction is found to be, in general, very small
and the single-particle potentials differ by at most 25 \% for
temperatures in the range from $0$ to $60$ MeV. The bulk
properties of infinite matter of baryons, either nuclear isospin
symmetric or a beta-stable composition which includes a non-zero
fraction of hyperons, are obtained. It is found that the presence
of hyperons can modify the thermodynamical properties of the
system in a non-negligible way.


\end{abstract}



\maketitle

\section{Introduction}
\label{sec:intro}

The Equation of State (EoS) is the essential ingredient for
understanding the behavior of nuclear matter under extreme
conditions of density and temperature. Recently it has received a
renewed interest due to the possibility of attaining such
conditions in relativistic heavy ion collisions at GSI, CERN and 
Brookhaven \cite{GSI,CERN,BNL}. In these
conditions, matter is expected to be at high densities and very
high temperatures. The EoS for larger densities and lower
temperatures is also important for the study of hot dense matter
in astrophysical conditions \cite{astro}. Supernova models involve
the gravitational collapse of the inner core of a massive star
followed by the explosive ejection of the overlying material and the
latter formation of a proto-neutron star. During
the collapse matter reaches densities beyond nuclear matter saturation
density ($\sim 0.17$ fm$^{-3}$) and temperatures of several tens of
MeV. Therefore, the description of matter in such conditions of density 
and temperature demands for the introduction of new degrees of freedom 
other than nucleons. In the case of cold neutron stars, {\it i.e.,} 
once neutrinos have been emitted and the temperature has dropped 
to negligible values, very different models have been
 used to describe their dense interior, ranging from quark matter to pion or
kaon condensates. Special attention has been focused on the
presence of hyperons ($\Lambda$ and $\Sigma^-$) in such media and
several calculations have been performed at $T=0$, and at finite
temperatures as well, within different phenomenological approaches
\cite{balb97,balb99} and relativistic mean field models
\cite{gl85,kn95,sc96,yang04}. Microscopic calculations have also been widely
applied to the study of cold neutron stars \cite{sc95,sc98,vi00,bal00,Vidana:2000ew,Vidana:ct}.
However, only few microscopic calculations of the EoS at finite
temperature are available \cite{le86,ba99,ba88,bo93,zuo03,ba04} and, to 
the best of our knowledge, the present work is the first microscopic calculation
of the EoS at finite temperatures including baryonic degrees of
freedom other than nucleons.

The microscopic approach employed in this work is based on the
Brueckner--Bethe--Goldstone (BBG) many-body theory. The basic
input of our calculations is the baryon-baryon interaction for the
complete baryon octet. For the nucleon-nucleon (NN) sector ({\it
i.e}, for neutrons $n$ and protons $p$) we use the realistic
Argonne V18 (Av18) potential \cite{argonne}. Nevertheless, since
it is well known that a Brueckner--Hartree--Fock (BHF) approach with
any realistic two body interaction does not yield satisfactory
saturation properties, we follow Refs.~\cite{le86,ba99} and supplement 
it with a Three-Body Force (TBF) which, after a suitable integration on
 the third particle, can be reduced to an effective two-body NN
interaction. The nucleonic bulk properties are then very well
described. On the other hand, for the hyperon-nucleon (YN) and
hyperon-hyperon (YY) sectors, we use the
Nijmegen NSC97e potential developed by Stoks and Rijken
\cite{st99}. In our microscopic treatment, temperature effects are
taken into account by modifying the momentum distributions and
Pauli blocking factors appropriately and solving the problem
self-consistently for a given density and temperature. In this
way, the effective in-medium interaction, the single-particle
properties and the bulk observables are modified by temperature.
This corresponds to the ``naive'' finite temperature BBG (NTBBG)
expansion referred to in Ref.~\cite{ba99}, where it was found to
be good enough for the low temperatures (up to $\sim 30$ MeV) explored
in that work.

The paper is organized in the following way. A brief
review of the Brueckner--Hartree--Fock (BHF)  approximation of the
BBG many-body theory at zero temperature extended to the hyperonic
sector is given in Sec.\ \ref{sec:zerotemp}. The extension to the
finite temperature case is presented in Sec.\
\ref{sec:finitetemp}. Section \ref{sec:results} is devoted to the
presentation and discussion of the results. In Sec.
\ref{sec:sp_prop} we discuss the results related to the
single-particle properties, while \ref{sec:bulk_prop} is devoted
to the bulk properties of the system. Finally, a short summary and
the main conclusions of this work are drawn in Sec.\
\ref{sec:conclusions}.

\section{Formalism}
\label{sec:formalism}


\subsection{Zero temperature BHF approximation}
\label{sec:zerotemp}

Our many-body scheme is based on the BHF approximation of the BBG theory
extended to the hyperonic sector \cite{sc95,sc98,vi00}. It starts with the
construction of all the baryon-baryon ({\it i.e.,} NN, YN and YY)
$G$-matrices, which
describe in an effective way the interactions between baryons in the
presence of a surrounding baryonic medium. They are formally obtained by
solving the well known Bethe--Goldstone equation, written schematically as
\begin{equation}
   G(\omega)_{B_1B_2,B_3B_4} =
   V_{B_1B_2,B_3B_4}+
  \displaystyle{ \sum_{B_5B_6}V_{B_1B_2,B_5B_6}
   \frac{Q_{B_5B_6}}{\omega-E_{B_5}-E_{B_6}+ i\eta}
   G(\omega)_{B_5B_6,B_3B_4} } \ .
   \label{eq:gmatrix} \end{equation} 
In the above expression the first
(last) two subindices indicate the initial (final)  two-baryon
states compatible with a given value $S$ of the strangeness,
namely NN for $S=0$, YN for $S=-1,-2$, and YY for $S=-2,-3,-4$, $V$ is 
the bare baryon-baryon interaction (Av18+TBF for NN, NSC97e for YN and YY),
 $Q_{B_5B_6}$ is the Pauli operator which prevents the
intermediate baryons $B_5$ and $B_6$ from being scattered to states below
their respective Fermi momenta, and $\omega$, the so-called starting
energy, corresponds to the sum of nonrelativistic single-particle energies
of the interacting baryons (see Ref.\ \cite{vi00} for computational
details).

The single-particle energy of a baryon $B_i$ is given by (we use units in
which $\hbar =1$, $c=1$)
\begin{equation}
E_{B_i}=M_{B_i}+\frac{k^2}{2M_{B_i}}+{\mathrm Re}[U_{B_i}(k)] \ ,
\label{eq:spe}
\end{equation}
where $M_{B_i}$ denotes the rest mass of the baryon and the real part of 
the single-particle potential $U_{B_i}(k)$
represents the averaged field ``felt'' by the baryon due to its
interaction with the other baryons of the medium. In the BHF approximation,
$U_{B_i}(k)$ is given by \begin{equation}
       U_{B_i}(k) =
        \sum_{B_j}\sum_{k'}n_{B_j}(k')
       \left\langle \vec{k}\vec{k'}\right |
       G_{B_iB_j,B_iB_j}(\omega=E_{B_i}+E_{B_j})
       \left | \vec{k}\vec{k'} \right\rangle  \ ,
\label{eq:upot}
\end{equation}
where
\begin{equation}
n_{B_j}(k) =
\left\{
 \begin{array}{ll} 1 , \mbox{if $k \leq k_{F_{B_j}}$} \\ 0 ,
\mbox{otherwise} \end{array} \right.  \label{eq:ocnumb} \end{equation} 
is
the corresponding occupation number of the species $B_j$, a sum over all
the different baryon species is performed and the matrix elements are
properly antisymmetrized when baryons $B_i$ and $B_j$ belong to the same
isomultiplet. We note here that the so-called continuous prescription has
been adopted for the single-particle potentials when solving the
Bethe--Goldstone--Equation at $T=0$. As shown by the authors of Refs.
\cite{so98,ba00}, the contribution to the energy per particle from
three-body clusters is diminished in this prescription. We note also that
the present calculations have been carried out using the Av18 potential
supplemented with a TBF in the NN sector and including the most recent
parametrization of the bare baryon-baryon potential for the YN and YY sectors,
 as defined by Stoks and Rijken in Ref.\ \cite{st99}. This
potential model, which aims at describing all interaction channels with
strangeness from $S=0$ to $S=4$, is based on SU(3) extensions of the
Nijmegen NN and YN potentials \cite{ri98}.

The calculations of all the other bulk properties of the cold system can then
be obtained from the total energy per particle, $E/A$, which is easily calculated
from the BBG expression:
\begin{eqnarray}
\frac{E}{A} &=& \frac{1}{A}\sum_{B_i}\sum_{k} n_{B_i}(k) 
\bigg\{\frac{k^2}{2M_{B_i}}  \nonumber \\ 
 && + \frac{1}{2} \sum_{B_j} \sum_{k'} n_{B_j}(k')
       \left\langle \vec{k}\vec{k'}\right |
       G_{B_iB_j,B_iB_j}(\omega=E_{B_i}+E_{B_j})
       \left | \vec{k}\vec{k'} \right\rangle \bigg\} \ ,
\label{eq:ea}
 \end{eqnarray}
 once a self-consistent solution of Eqs.\ (\ref{eq:gmatrix})--({\ref{eq:upot}) is achieved.


\subsection{Finite temperature effects} \label{sec:finitetemp}

The many-body problem at finite temperature have been considered by several
authors within different approaches, such as the finite temperature Green's
function method \cite{fe71}, thermo-field dynamics \cite{he95}, or the
Bloch--De Dominicis (BD) diagrammatic expansion \cite{bl58}. The latter,
developed soon after the Brueckner theory, represents the ``natural''
extension to finite temperature of the BBG expansion, to which it leads in
the zero temperature limit. Baldo and Ferreira \cite{ba99} showed that the
dominant terms in the BD expansion were those that correspond to the zero
temperature BBG diagrams but introducing the temperature in
the Fermi-Dirac distributions which now read
\begin{equation}
f_{B_i}(k,T)=\frac{1}{1+exp([E_{B_i}(k,T)-\widetilde{\mu}_{B_i}]/T)}
\ , \label{eq:fd}
\end{equation}
where $\widetilde{\mu}_{B_i}$ is the chemical potential of the
baryon species $B_i$.

Therefore, at the BHF level, finite temperature effects can be
introduced in a very good approximation just changing in the
Bethe--Goldstone equation: (i) the zero temperature Pauli operator
$Q_{B_5B_6}=(1-n_{B_5})(1-n_{B_6})$ by the corresponding finite
temperature one $Q_{B_5B_6}(T)=(1-f_{B_5})(1-f_{B_6})$, and (ii)
the single-particle energies $E_{B_i}$ by the temperature
dependent ones $E_{B_i}(T)$ obtained from Eqs.\ (\ref{eq:spe}),
(\ref{eq:upot}) by replacing $n_{B_i}(k)$ by $f_{B_i}(k,T)$. These
approximations, which we will suppose valid in the range of
densities and temperatures considered here, were referred in
\cite{ba99} as the ``naive'' finite temperature (NTBBG) expansion.

In this case, however, the self-consistent process implies that, together
with the Bethe--Goldstone equation and the the single-particle potentials,
the chemical potential of each baryon species, $\widetilde{\mu}_{B_i}$,
must be extracted at each step of the iterative process from the
normalization condition
\begin{equation}
\rho_{B_i}=\sum_{k} f_{B_i}(k,T)
\ . \label{eq:cp}
\end{equation}
This is an implicit equation for each one
of the chemical potentials, which can be solved numerically. Note that, now,
also the Bethe--Goldstone equation
and single-particle potentials depend implicitly on the chemical
potentials.

Once a self-consistent solution is obtained, the total free energy per
particle is determined by
\begin{equation}
\frac{F}{A}=\frac{E}{A}-T\frac{S}{A} \ , \label{eq:free_ener}
\end{equation}
where $E/A$ is evaluated from Eq.\ (\ref{eq:ea}) replacing
$n_{B_i}(k)$ by $f_{B_i}(k,T)$ and the total entropy per particle,
$S/A$, is calculated using the following mean-field expression:
\begin{equation}
\frac{S}{A}=-\frac{1}{A}\sum_{B_i}\sum_{k}[f_{B_i}(k,T)\mbox{ln}(f_{B_i}(k,T))+
(1-f_{B_i}(k,T))\mbox{ln}(1-f_{B_i}(k,T))] \ . \label{eq:entrop}
\end{equation}

We note that in a thermodynamically consistent (conserving)
approach, the value of the chemical potential
$\widetilde{\mu}_{B_i}$ obtained from the normalization condition
(\ref{eq:fd}) should coincide with the thermodynamical definition
\begin{equation}
\mu_{B_i} = \frac{\partial F}{\partial N_{B_i}} \ . \label{eq:nu}
\end{equation}
This is not the case of the BHF approximation employed here,
neither at zero nor at finite temperature, as it is well known.
Another manifestation of the non-conserving character of the BHF
approach is the violation of the Hugenholtz-van Hove theorem.
Further remarks on this problem and how the validity of the
Hugenholtz-van Hove theorem can be restored will be given in the
next section.


\section{Results}
\label{sec:results}

\subsection{Single-particle properties}
\label{sec:sp_prop}

We start this section by discussing the behavior of the momentum
distribution and the Pauli operator with temperature, both being crucial
ingredients in determining the dependence with temperature of all the other
physical quantities. The momentum distribution in nuclear matter at 
experimental saturation density $\rho_0=0.17$ fm$^{-3}$ is shown in 
Fig.~\ref{fig:mom} for various temperatures: $T=0$ (solid line), 
20 (short-dashed lines), 40 (long-dashed lines) and 60 MeV (dot-dashed lines).
 The thick lines are momentum distributions obtained with a purely kinetic 
energy spectrum and will be denoted by $f_{\rm free}(k,T)$ henceforth. The 
thin lines, which will be denoted as $f(k,T)$, contain the effect of a 
single-particle potential. It may appear surprising that the depletion of 
low momentum states for a non-interacting system at finite temperature is 
larger than that for the interacting system at the same temperature, 
{\it i.e.,} $f_{\rm free}(0,T) < f(0,T)$. This is a direct consequence of 
the momentum dependence of the spectrum and it is the behavior to be 
expected for nucleons in nuclear matter, which have a steeper spectrum
than the purely kinetic energy one, hence making the single-particle
excitations more costly than in the free system.

\begin{figure}[htb] \centering
\includegraphics[width=9cm]{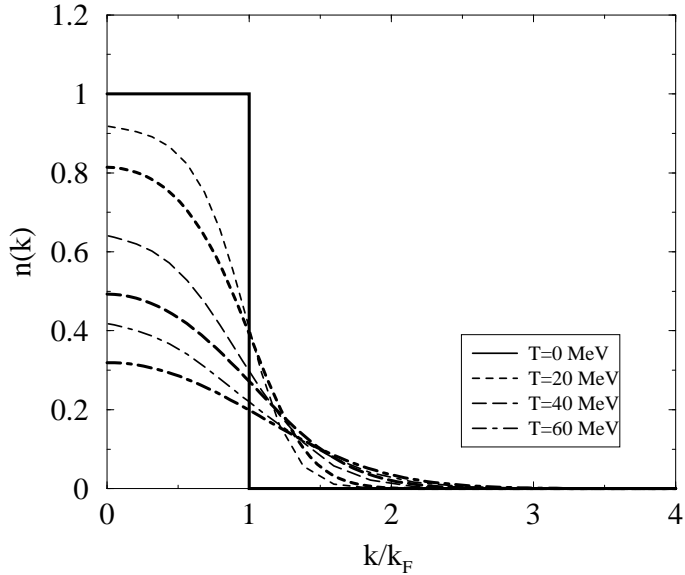} 
\caption{Nucleon momentum
distributions at $\rho_0$ for various temperatures: $T=0$, $20$, $40$ and $60$ MeV. The thick lines have been obtained with a purely kinetic energy
spectrum, while the thin lines contain the effect of a single-particle
potential.} \label{fig:mom} \end{figure}

The corresponding angle averaged Pauli blocking factor between two
different baryons $B_1$ and $B_2$ reads:
\begin{equation}
\bar{Q}_{B_1 B_2}(k,K)=\frac{1}{2}\int_{-1}^1 d(\cos \theta) 
[1-f_{B_1}(\mid \alpha \vec{K} + \vec{k} \mid,T)]
[1-f_{B_2}(\mid \beta \vec{K} - \vec{k}\mid,T)] \ ,
\end{equation}
where $\alpha=\frac{M_1}{M_1+M_2}$ and
$\beta=\frac{M_2}{M_1+M_2}$. In Fig.~\ref{fig:pauli}, we show the
nucleon-nucleon Pauli blocking factor as a function of relative
momentum $k$ and for two different total momenta, $K=0$ (left
panel) and $K=3 k_F$ (right panel), calculated at the previously
stated density and temperatures. One can see that the sharp
behaviors which characterized the Pauli blocking at zero
temperature  smear out considerably with temperature. In addition,
 the figure also illustrates the loss of Pauli blocking effects
as temperature increases. Due to this gain in phase space, medium
effects weaken with increasing temperature.

\begin{figure}[htp] \centering
\includegraphics[width=8cm]{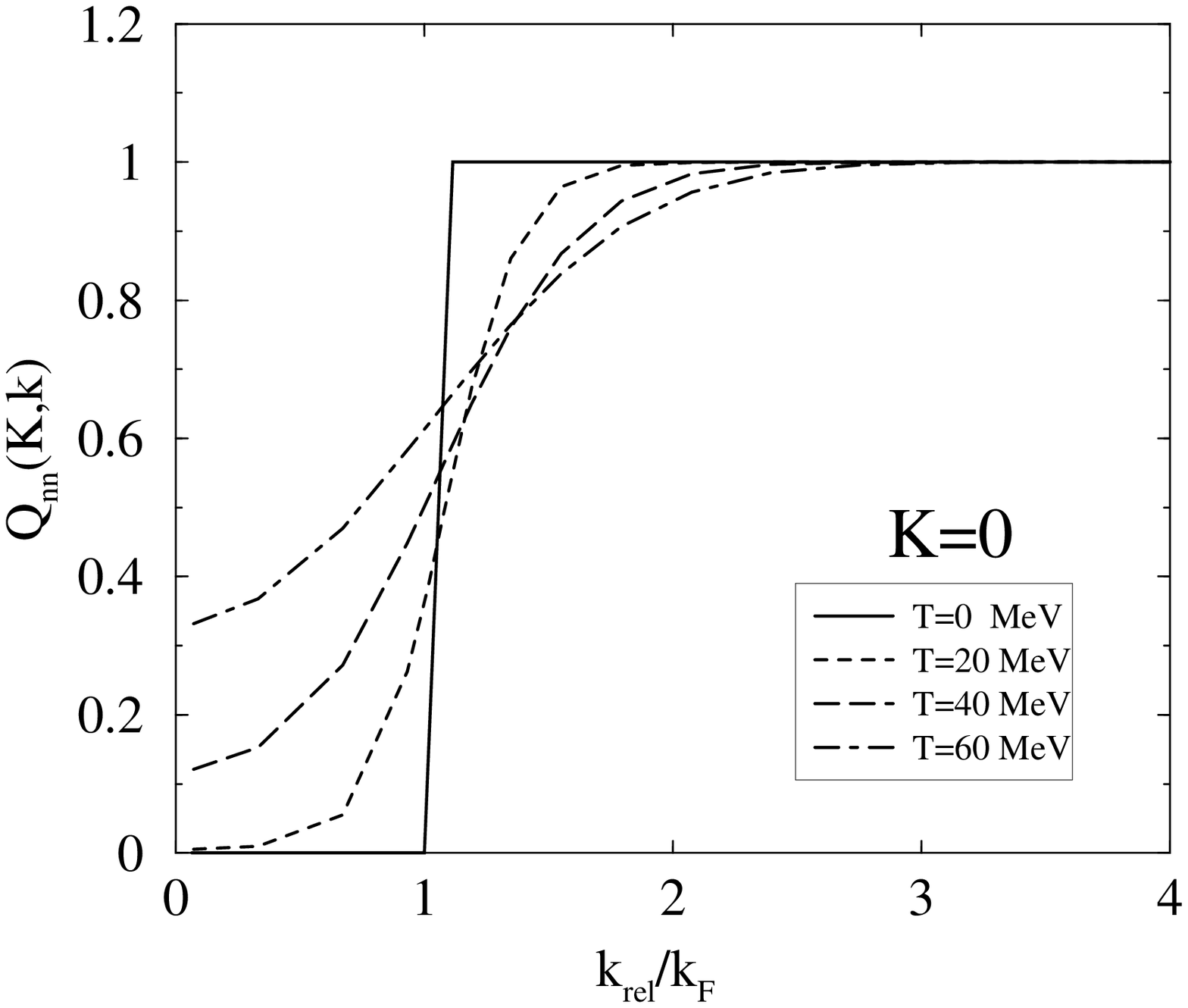}
\includegraphics[width=8cm]{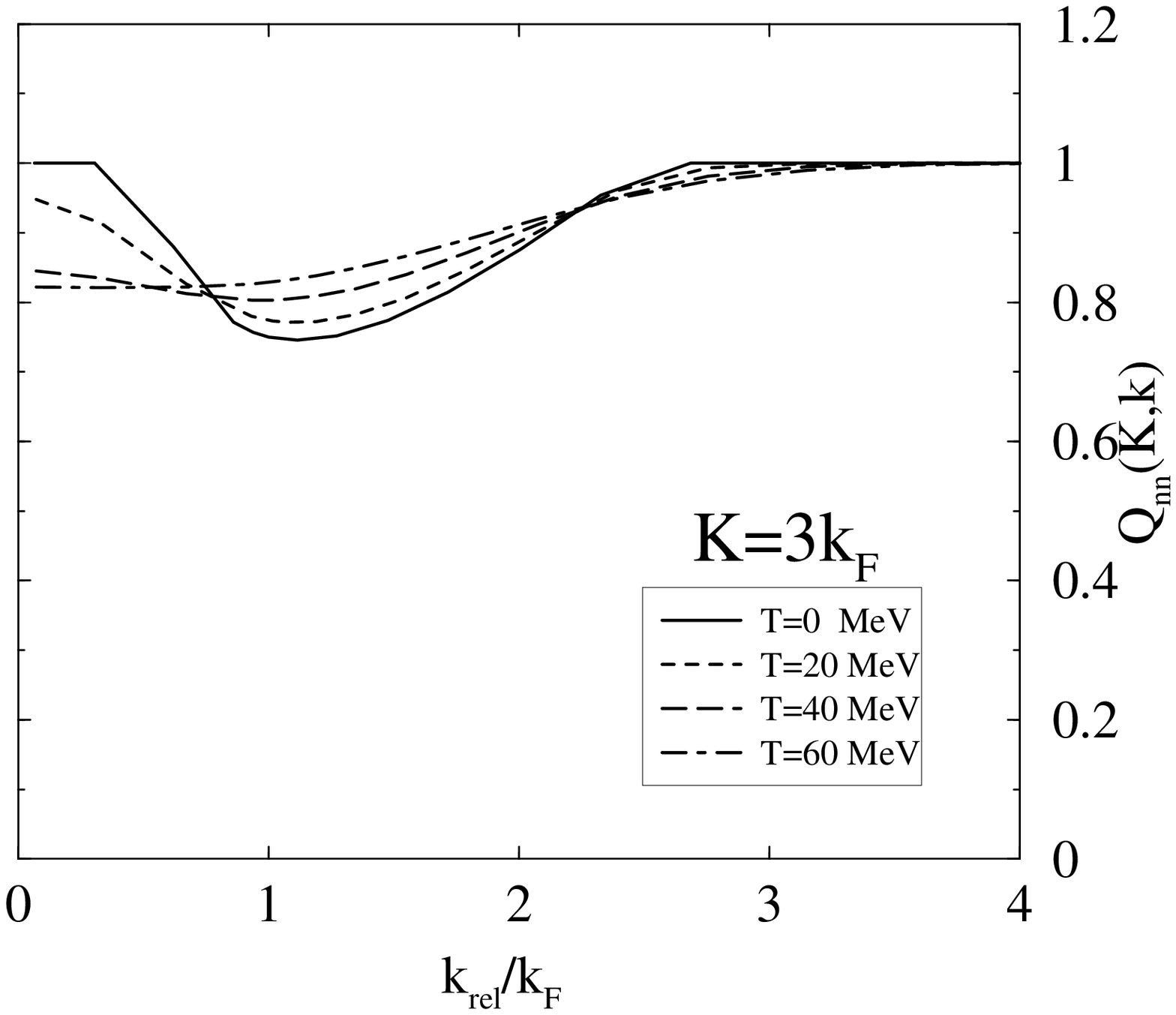} 
\caption{Pauli blocking factor at $\rho_0$ for various temperatures: $T=0$, 
$20$, $40$ and $60$ MeV. The center of mass momentum are $K=0$ fm$^{-1}$ (left panel) and $K=3k_F$ fm$^{-1}$ (right panel).} 
\label{fig:pauli} \end{figure}

We next discuss the behavior of the NN effective in-medium
interaction with temperature. The $^1S_0$ G-matrix element is
shown in Fig.~\ref{fig:1s0} as a function of the on-shell relative
momemtum for a total center-of-mass momentum $K=0$. While the real
part shows little dependence with temperature, the imaginary part
reflects in a more explicit way the behavior of the Pauli blocking
factor. At finite temperature, a non-zero imaginary part appears 
below $k_F$ due to the depletion in the occupation of single-particle 
momentum states. As temperature increases, the size of the imaginary part
becomes larger in this region. This is needed since, at very high T, 
the G-matrix should approach the free T-matrix elements which have a narrow
structure in the low momentum region due to the almost existence of a bound
state in the $^1S_0$ partial wave. The G-matrix for the $^3S_1$ partial 
wave is shown in Fig.~\ref{fig:3s1}. The larger dependence on temperature
observed for this partial wave is a direct consequence of the existence of
the deuteron pole in the $^3S_1$ T-matrix.

\begin{figure}[htp] \centering \includegraphics[width=9cm]{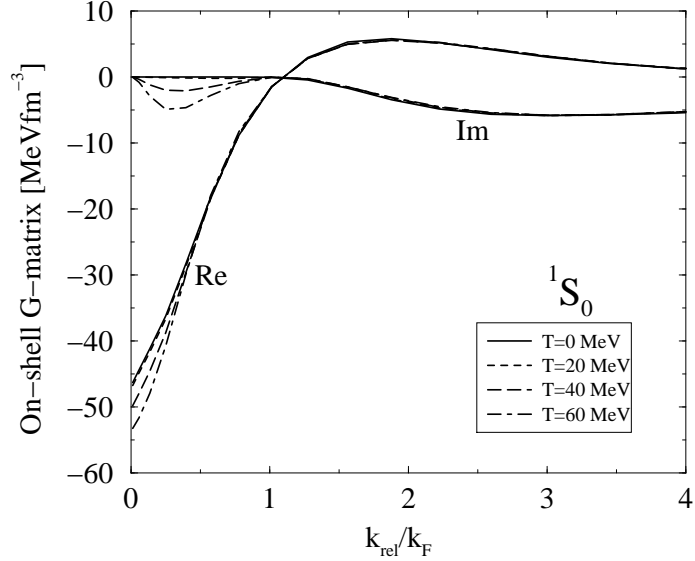}
\caption{$^1S_0$ on-shell G-matrix elements as a function of 
relative momentum at $\rho_0$ for various temperatures: $T=0$, 
$20$, $40$ and $60$ MeV.} 
\label{fig:1s0} \end{figure}

\begin{figure}[htp] \centering
\includegraphics[width=9cm]{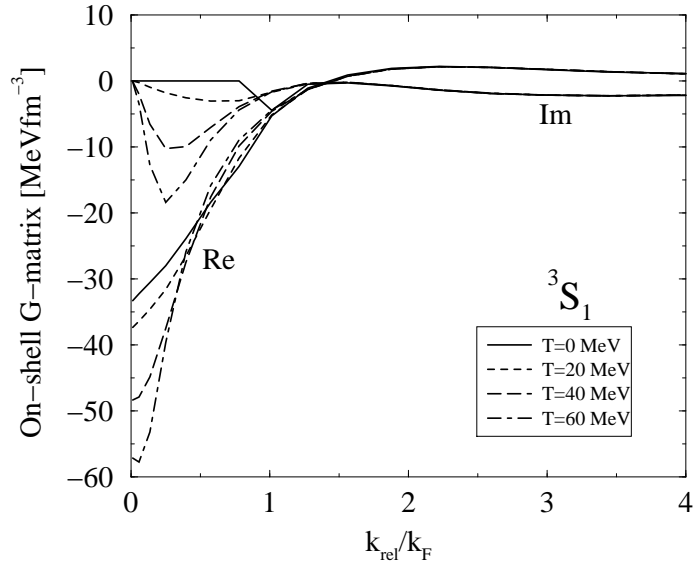} 
\caption{The same as Fig.~\protect\ref{fig:1s0} for the $^3S_1$ 
partial wave.}
\label{fig:3s1} \end{figure}

Another important quantity, which is directly related to the
propagation properties of a baryon $B_1$ inside a medium of other
baryons, is the single-particle potential $U_{B_1}(k)$. In
Fig.~\ref{fig:udksym} we display the full momentum dependence of
these single-particle potentials for all the octet baryons ($N$,
$\Lambda$, $\Sigma$ and $\Xi$) in pure symmetric nuclear matter at
normal density $\rho_0$ for various temperatures, $T=0$ (solid lines), 
$20$ (short-dashed lines), $40$ (long-dashed lines) and $60$ MeV 
(dot-dashed lines). We will first
discuss the momentum dependence of the $T=0$ results. The
single-particle potentials at $T=0$ correspond to the curves which
have a stronger structure, due to to more abrupt Pauli blocking
and threshold effects. For nucleons, the real part of the
single-particle potential shows a cusp around the Fermi momentum,
which is a reflection of the in-medium deuteron structure of the
$^3S_1$ G-matrix elements. For the $\Lambda$ and $\Sigma$ ($\Xi$)
hyperons, the $T=0$ potential is the one that gives the more
attractive (repulsive) potential at the origin. In the particular
case of the $\Lambda$ hyperons, the real part of $U_{\Lambda}$
shows a structure around 3 fm$^{-1}$ related to the opening of the
$\Sigma N$ threshold. This is confirmed by the appearance of an
additional source of imaginary part in $U_{\Lambda}$ around this
momentum. The imaginary part of the $\Sigma$ potential,
$U_{\Sigma}$, is already finite at zero momentum due to its decay
through the processes $\Sigma N\to \Lambda N$. Finally, the
structures found for the $\Xi$ hyperons around 2 and 3 fm$^{-1}$
reflect the cusps observed in the $\Xi N$ cross sections at
energies corresponding to the opening up of the $\Sigma\Lambda$
and $\Sigma\Sigma$ channels \cite{st99}, respectively.

\begin{figure}[htp] \centering \includegraphics[width=9cm]{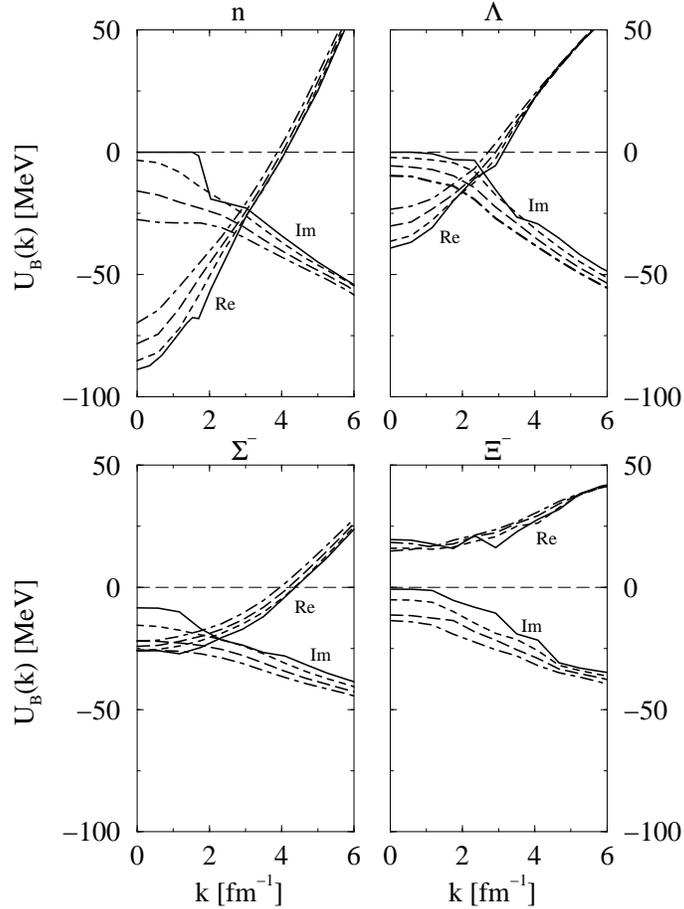}
\caption{Single-particle potential of the octet baryons at $\rho_0$ as
 a function of momentum for various temperatures: $T=0$ (solid lines), 
$20$ (short-dashed lines), $40$ (long-dashed lines) and $60$ MeV 
(dot-dashed lines).} 
\label{fig:udksym}
\end{figure}

With respect to the temperature dependence, we see that the
single-particle potentials change gradually as temperature
increases. For the momenta explored, the imaginary part of the
single-particle potential for all baryons increase in size with
temperature, as a consequence of the increase of phase space in
the low momentum region. The temperature effects on the imaginary
parts are more important near the origin. When going from $T=0$ to
$T=60$ MeV, their increase is of the order of $25$ MeV for
nucleons, $10$ MeV for $\Lambda$'s and $15$ MeV for the $\Sigma$
and the $\Xi$ hyperons. In the nucleon sector, the attractive real
part of the single-particle potential becomes more and more
repulsive as temperature increases (an effect which, near the
origin, involves a $20 \%$ correction at $T=60$ MeV with respect
to the $T=0$ result). This may seem contradictory with the small
gain in attraction of the NN G-matrix elements at $T=60$ MeV,
shown in Figs.~\ref{fig:1s0} and
 \ref{fig:3s1}. Note, however, that the finite temperature momentum distribution also
allows for the contribution of higher relative momentum states to
$U(k)$. Since in this region the effective interaction is less
attractive, the net effect on $U(k)$ is a loss in the attraction
as temperature increases. This is in accordance with what was
observed in the pioneering work of Ref.~\cite{le86}. The real part
of the $\Lambda$ and $\Sigma$ single-particle potentials also show
the same trend and, in the origin, they increase a $30 \%$ and a
$5 \%$ respectively when rising the temperature from $0$ to $60$
MeV. However, in contrast to the other baryons, the real part of
the $\Xi$ single-particle potential is repulsive and, at zero
momentum, the potential experiences an attractive gain of about
$2$ MeV as temperature is increased.

\begin{figure}[htp] \centering
\includegraphics[width=12cm]{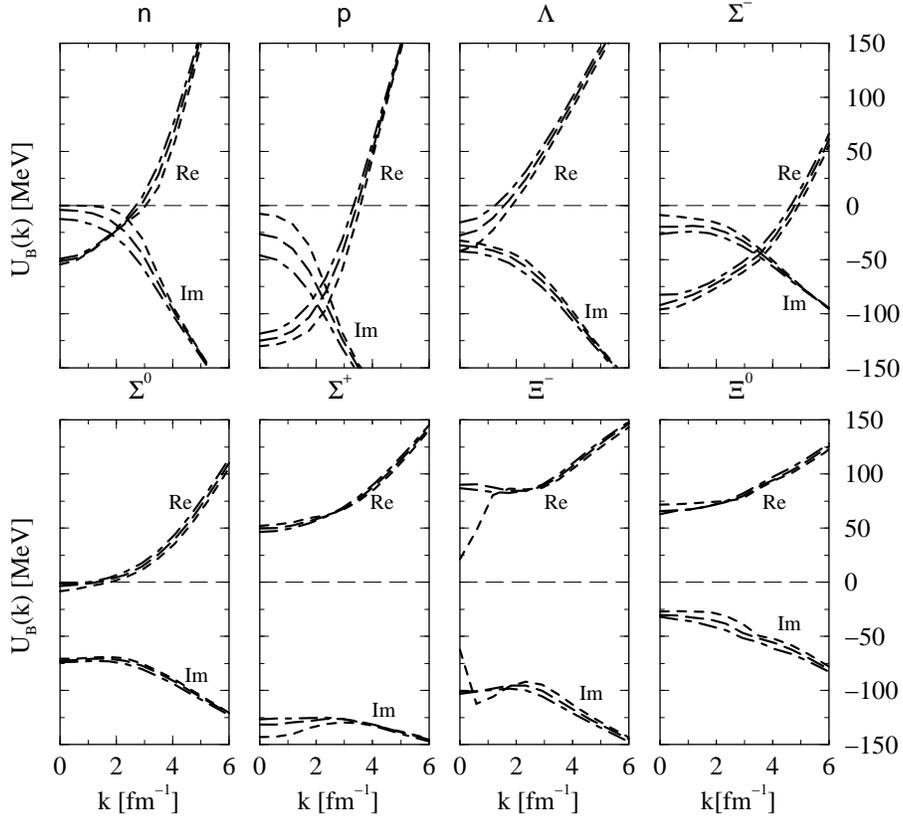}
\caption{Single-particle potential of the octet baryons as a function of
momentum in baryonic matter at $\rho=0.5$ fm$^{-1}$ composed by
80\% neutrons, 10\% protons and 10\% $\Sigma^-$ hyperons, for
various temperatures: $20$ (short-dashed lines), $40$ (long-dashed lines) 
and 60 MeV (dot-dashed lines).} \label{fig:udk05}
\end{figure}

Next, in Fig.~\ref{fig:udk05}, we show the finite temperature
single-particle potentials of all the baryons of the ground-state octet in a
baryonic matter of total density $\rho=0.5$ fm$^{-1}$. At this
density, one expects matter to contain additional degrees of
freedom, such as hyperons. We have chosen a composition of
neutrons, protons and $\Sigma^-$ hyperons in a proportion of 80\%,
10\% and 10\%, respectively. This corresponds to a representative
example of the composition of a cold neutron star in
beta-equilibrium \cite{Vidana:2000ew}. This case is interesting
because the finite fraction of hyperons makes the properties of
matter depend not only on the $NN$ interaction, but also on both
the $YN$ and $YY$ in-medium effective interactions.

Regarding the momentum dependence, all the single-particle
potentials have a steeper growth with respect to the previous case
(note the different vertical scale in Fig.~\ref{fig:udk05}
compared to that in Fig.~\ref{fig:udksym}), in accordance with the
fact that the system is at a higher density. The most salient
feature is, however, the different behavior of the single-particle
potential for the various members of the same isospin multiplet
due to the strong isospin asymmetry in the composition of matter.
As expected, protons become more attractive than neutrons due to
the excess of $np$ interacting pairs in the strongly attractive
$^3S_1-^3D_1$ $I=0$ channel. The differences found for the
$\Sigma$ hyperons are also interesting: the $\Sigma^-$
single-particle potential receives a large contribution from the
interaction of I=3/2 $\Sigma^- n$ pairs, which is strongly
attractive for the NSC97e model used in this work. Conversely, the
repulsive $\Sigma^+$ potential is basically built from the
interaction of $\Sigma^+ n$ pairs which, apart from the attractive
$I=3/2$ component, also receives contributions from the very
repulsive $I=1/2$ component of the $\Sigma N$ NSC97e interaction.
Finally, the $\Sigma^0$ is not affected by the asymmetry between
neutrons and protons and its potential is mildly attractive. The
single-particle potential of the $\Xi^-$ particle at $T=20$ MeV
shows a strong structure at low momentum, which is also related to
the opening of the $\Lambda \Sigma^-$ and $\Sigma^0\Sigma^-$
channels as discussed in the previous figure for symmetric nuclear
matter at $\rho=\rho_0$, but occurring now at lower energy due to
the more attractive single-particle potentials felt by the
$\Lambda$ and the $\Sigma^-$ at this higher density. This effect
is magnified by the relatively larger amount of $\Xi^- n$
interacting pairs in this highly asymmetric baryonic matter
containing a neutron fraction of $80 \%$.

With respect to the temperature dependence, for all the baryon
species studied, it is found that the attractive single-particle
potentials become slightly less attractive and the repulsive
potentials become also less repulsive with increasing temperature.
In other words, the real parts of the single-particle potential
loose strength as temperature increases. It is also interesting to
note that, in this case, the single-particle potential of the
neutron (which is the most abundant species) is only slightly
modified with temperature despite the higher density. On the other
hand, protons, which are much less abundant, are more affected by
temperature, especially the imaginary part of the single-particle
potential, which increases by almost $40$ MeV when the temperature
changes from $20$ to $60$ MeV. The real part of the $\Lambda$
single-particle potential at the origin also increases
considerably, by almost $30$ MeV. For the other particles, except
for the already commented case of the $\Xi^-$, the changes in the
single-particle potential (both real and imaginary parts) amount
to 10--15 MeV.

\begin{figure}[htp]\centering \includegraphics[width=9cm]{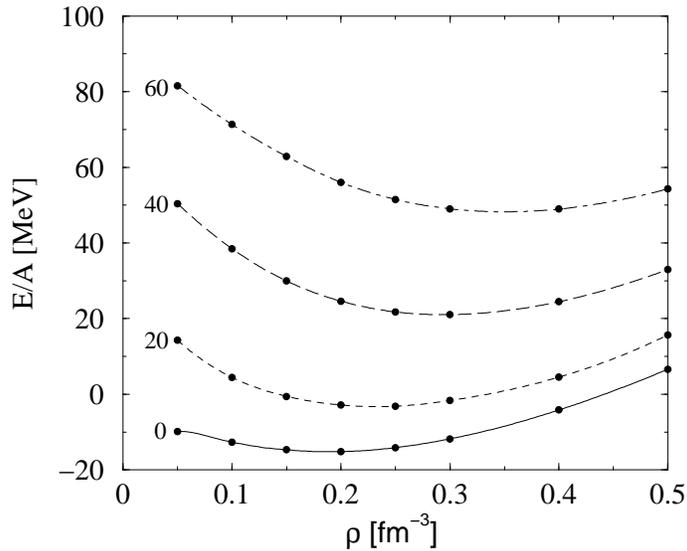} 
\caption{ Energy per nucleon of symmetric nuclear matter as a function 
of density for various temperatures: 0, 20, 40 and 60 MeV.}
\label{fig:ener_sym}
\end{figure}

\begin{figure}[htp] \centering \includegraphics[width=9cm]{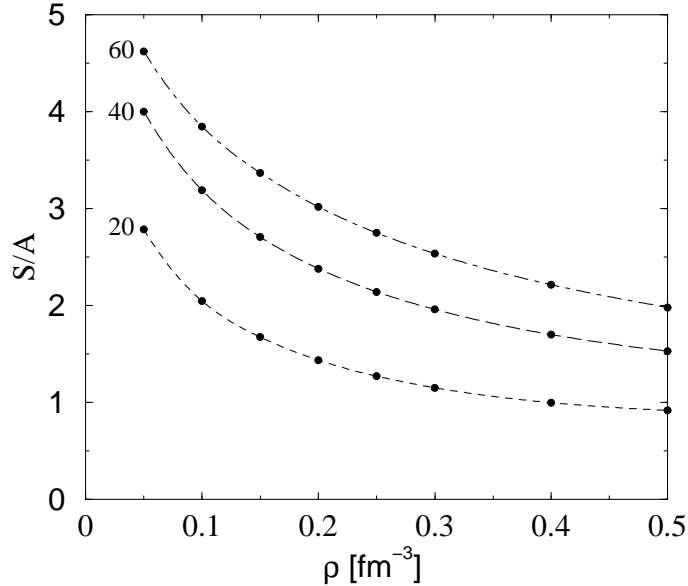} 
\caption{Entropy per nucleon of symmetric nuclear matter as a function 
of density for various temperatures: 20, 40 and 60 MeV.}
\label{fig:entro_sym}
\end{figure}

\subsection{Bulk properties}
\label{sec:bulk_prop}

From the astrophysical point of view, the main interest lies in
the thermodynamical properties of the extended system of baryons
at finite temperature. These are the properties that we will
discuss from now on. In Fig.~\ref{fig:ener_sym}, for instance, we
report our results for the internal energy per nucleon, $E/A$,  of
symmetric nuclear matter as a function of density for various
temperatures. At $T=0$ we can reproduce the saturation properties
thanks to the implementation of an additional TBF on top of the
Av18 interaction. We note that $E/A$ increases considerably with
temperature for all densities, basically due to the increase of
kinetic energy. The interacting part of the energy becomes
slightly less attractive as temperature rises. For instance, at
$T=60$ MeV the interaction energy is only about $15$ MeV higher
than in the $T=0$ case in the whole range of densities explored,
while the non-interacting part of the internal energy increases an
amount that varies between $50$ and $90$ MeV in the same density
range. We also note that the internal energy at $T=60$ MeV and
$\rho=0.05$ fm$^{-3}$ has practically reached the non-degenerate
classical limit ($\varepsilon_F/T = 0.14 << 1$) where $E/A$ would
amount to $\frac{3}{2}T = 90$ MeV, the reason being that in this
case the potential energy is negligible with respect to the
kinetic contribution.

\begin{figure}[htp]\centering \includegraphics[width=9cm]{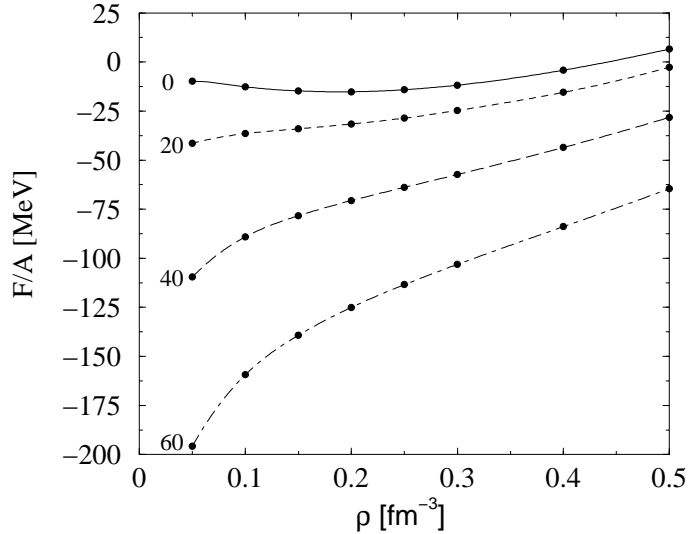} 
\caption{Free energy per nucleon of symmetric nuclear matter as a function
of density for various temperatures: 0, 20, 40 and 60 MeV.}
\label{fig:free_sym}
\end{figure}

The entropy per particle, as calculated from the mean field
approximation expression of Eq.~(\ref{eq:entrop}), is displayed in
Fig.~\ref{fig:entro_sym}. The values obtained at $T=20$ MeV are
similar to those in the work of Baldo and Ferreira \cite{ba99},
where results for temperatures up to $T=28$ MeV are reported. As
expected,  the entropy increases with $T$. We also observe that, at a given
temperature, the entropy is always largest at the lowest density, where one is
closer to the classical limit, and decreases smoothly with increasing density as
the system evolves to the degenerate regime.

In Fig.~\ref{fig:free_sym} we show our results for the free energy per nucleon
obtained from Eq.~(\ref{eq:free_ener}). For temperatures larger than 20 MeV,
and in all the range of densities explored, the entropic negative contribution,
$-T S$, dominates over the internal energy one, $E$. This is evidently more
pronounced at low densities, when one is closer to the classical limit where the
entropy becomes very large.
The limited amount of temperatures explored in this work prevents us from giving
a precise value for the liquid-gas transition temperature $T_C$. However, the
similarity of our results around $T=20$ MeV with those reported in
Ref.~\cite{ba99} allows us to conclude that our value of $T_C$ is
close to 15-20 MeV, similarly to other BHF approaches using different NN
interactions \cite{ba99,ba88,bo93,zuo03,ba04}.

We recall that in the present study we are working in the so-called
NTBBG approach, which corresponds to truncating the two-body
correlation contribution of the BD diagrammatic expansion of the
grand-canonical potential to first order in the two-body
scattering matrix. This procedure, valid for low temperatures, is
formally analogous to the Brueckner--Hartree--Fock binding potential
of the zero temperature case but replacing the Fermi step
functions by the finite temperature Fermi-Dirac distributions. An
explicit calculation allowed the authors of Ref.~\cite{ba99} to
conclude that, in the temperature and density range explored in
that work, the higher order terms neglected in the finite
temperature expansion represented at most a few percent of the
leading term, so they could be safely neglected. Since we are
considering here values of temperature of up to about twice the
highest temperature explored in Ref.~\cite{ba99}, we expect
slightly larger errors but not more than 10\% of the interaction
terms, {\it i.e.,} the higher order contributions would increase the free
energy per particle by less than $5$ MeV at $T=60$ MeV.

\begin{figure}[htp] \centering \includegraphics[width=9cm]{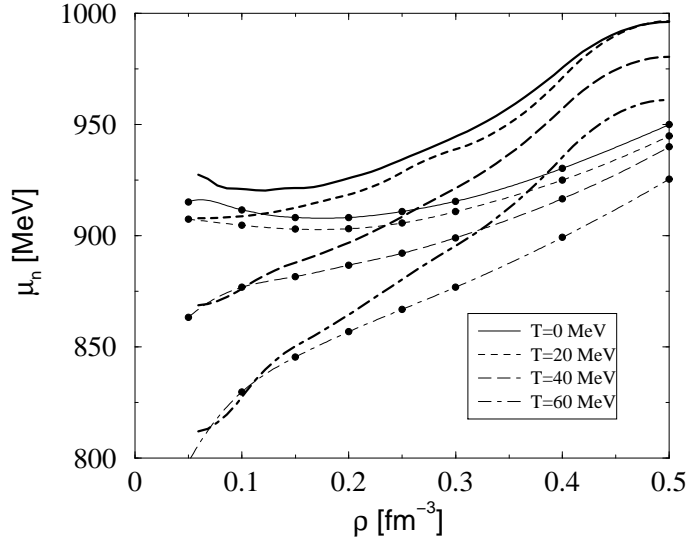} 
\caption{ Nucleon chemical potential for various temperatures as a function
of the density. The solid circles correspond to the chemical potential $\tilde{\mu}$,
obtained from the normalization condition of Eq.~(\protect\ref{eq:fd}), 
while the thick lines correspond to $\mu$, the chemical potential 
obtained from the derivative of the free energy.} 
\label{fig:chemn_num}
\end{figure}

The chemical potential of the nucleons as a function of density is
shown in Fig.~\ref{fig:chemn_num} for various temperatures. The
thick lines correspond to the chemical potential obtained from the
derivative of the free energy, Eq.~(\ref{eq:nu}), while the thin
lines which interpolate the calculated points represent the
chemical potential extracted from the density according to the
normalization condition shown in Eq.~(\ref{eq:fd}). It is well
known that the discrepancies observed from
 both procedures reflect a violation of the Hugenholtz-Van
Hove theorem \cite{hugen58} characteristic of a non-conserving
approximation. The reason lies in the fact that, in the BHF method
adopted here, the so-called ``rearrangement" terms are absent.
Therefore, the single-particle spectrum is, in general, less
repulsive and, as a consequence, the chemical potential from the
normalization condition is smaller than that obtained using the
derivative of the free energy. From Fig.~\ref{fig:chemn_num} we
can see that rearrangement effects are more important at higher
densities, of the order of $50$ MeV at the highest density in the
$T=0$ case. Their size reduce with increasing temperature and at
$T=60$ MeV they amount to at most $35$ MeV. Such a decrease is
related to the reduction of medium effects with increasing
temperature and, as a consequence, the density dependence of the
effective interaction, which originates the rearrangement terms,
becomes also less important.

One may restore the fulfillment of the Hugenholtz-Van Hove
theorem by adding the difference of chemical potentials to the BHF
single-particle potential energy $U(k)$. This shift will affect the
G-matrix and, therefore, the Brueckner energy of Eq. \ref{eq:ea}. However,
one should also consistently add the contribution of the bubble and 
the potential insertion diagrams which no longer cancel. As pointed out
in Ref.~\cite{ba99} this correction compensates quite accurately the 
modification of the Brueckner energy at $T=0$ and, 
 to the extent that the NTBBG is a reasonable
approximation at low temperatures, it does not affect our finite T
results either. Therefore, our thermodynamical quantities,
together with a chemical potential extracted from the derivative
of the free energy, can be considered as those of a conserving
approximation.

Finally, we present in Figs.~\ref{fig:ener_05} to
\ref{fig:free_05} our results for the thermodynamic quantities in
the case of baryonic matter with a neutron fraction $x_n=0.8$, a
proton fraction $x_p=0.1$ and a $\Sigma^-$ hyperon fraction
$x_{\Sigma}=0.1$. Although strictly speaking this composition is
representative of beta stable hyperonic matter at $T=0$ and
$\rho=0.5$ fm$^{-3}$, we have kept it in a wide range of densities
and temperatures in order to see the relevance of having a finite
fraction of hyperons on the bulk properties of this hypothetical
hadronic matter. The thin lines display the nucleonic-only
contribution normalized per baryon, so that we can clearly
distinguish the hyperon effects on the thermodynamical
properties.}

Our results for the internal energy are shown in
Fig.~\ref{fig:ener_05}. We first discuss the behavior of the
nucleonic contributions. With this composition, the nucleonic
contribution corresponds essentially to that of neutron matter,
modified by a 10\% proton fraction which induces the presence of
neutron-proton pairs interacting through the very attractive
isospin zero NN force. The smooth increase of $E$ with density at
$T=0$ MeV is a result of the combination of the increasing free
Fermi gas power law of $\rho^{2/3}$ for $E_{\rm kin}$ with a
smooth attractive behavior for $E_{\rm pot}$, which has a
saturating minimum of $-25$ MeV at $\rho=0.27$ fm$^{-3}$. As
temperature increases, $E_{\rm kin}/A$ increases very rapidly,
especially at low densities, where it approaches the classical
limit of $\frac{3}{2} T x_N$. The resulting $E/A$ is then
dominated by the non-interacting contribution with some slight
modulation from the interaction terms, which increase by about 10
MeV when going from $T=0$ to $T=60$ MeV, still maintaining the
saturating shape. We now discuss the differences between the thick
and thin lines which are due to the $\Sigma^-$ hyperonic
contributions. At zero-temperature, the repulsive hyperonic free
Fermi gas contribution is very small compared to the interaction
contributions, which come essentially from $\Sigma^- n$ pairs and,
to a minor extent,
 from $\Sigma^-\Sigma^-$  pairs. Both of them interact
attractively for the Nijmegen potential used here \cite{st99}.
Comparing the total internal energy at zero temperature with the
nucleonic contribution, one also observes that, as expected for
the interaction terms to the energy per baryon of a very diluted
system, the size of the attractive hyperonic contribution
increases practically linearly with density up to a value of $-6$
MeV at $\rho=0.5$ fm$^{-3}$. At finite temperatures and because of
the very low hyperonic density, the kinetic hyperonic component,
$E_{kin}/A$, acquires rapidly the classical value of $\frac{3}{2}
T x_Y$ for all densities, while the interaction terms depend
little with temperature. As a consequence, the hyperonic kinetic
energy overcomes the potential energy and the total hyperonic
contribution to $E/A$ switches from being attractive to being
repulsive when the temperature increases from zero to $60$ MeV.
Note, however, that the hyperonic interacting terms still play a
role at $T=60$ MeV and should not be neglected. At twice normal
nuclear matter density and $T=60$ MeV, they represent about half
of the hyperonic kinetic energy contribution.

\begin{figure}[htp] \centering \includegraphics[width=9cm]{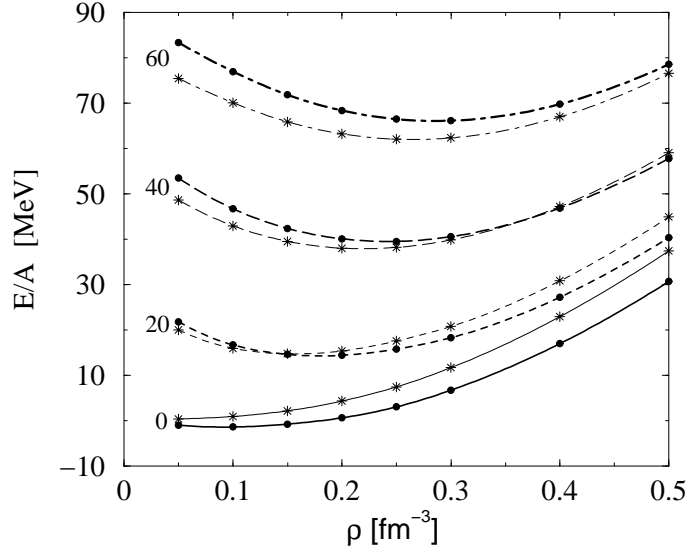} 
\caption{Internal energy per baryon as a function of density of baryonic
matter composed by 80\% neutrons, 10\% protons and 10\% $\Sigma^-$
hyperons, for various temperatures: 0, 20, 40 and 60 MeV.
The total (solid circles) and nucleonic-only (stars) contributions 
to the internal energy per baryon are represented by the thick and thin 
lines, respectively.}
 \label{fig:ener_05}
\end{figure}

The entropy per baryon is represented in Fig.~\ref{fig:entro_05}. As expected,
the total entropy per baryon decreases smoothly with increasing density as one
moves away from the classical limit. By comparing the total (thick lines) with
the nucleonic-only (thin lines) contributions we observe a strong influence of
the hyperons in the final value of the entropy, in spite of them representing
 only a 10\% fraction of the total number of particles.
The reason lies precisely in the fact that the hyperons in this system form
a very diluted gas which behaves almost classically, hence having
large entropy values.

\begin{figure}[htp] \centering \includegraphics[width=9cm]{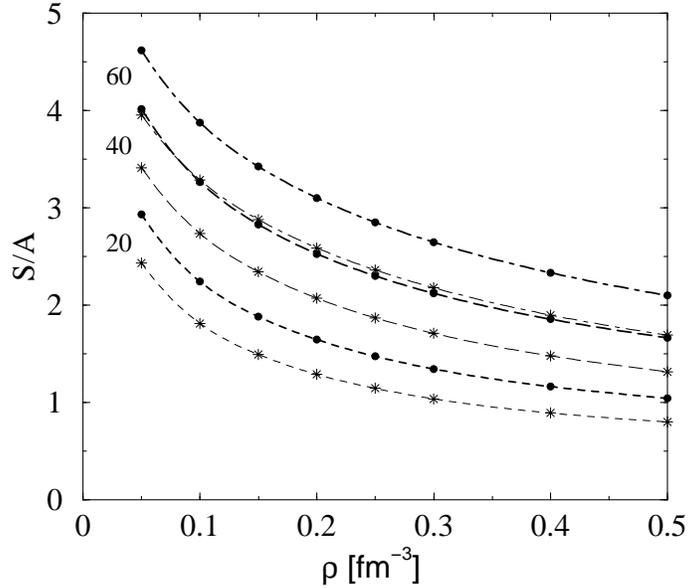}
\caption{Entropy per nucleon as a function of density of baryonic matter
composed by 80\% neutrons, 10\% protons and 10\% $\Sigma^-$
hyperons, for various temperatures: 20, 40 and 60 MeV. The total 
(solid circles) and nucleonic-only (stars) contributions to the entropy 
per baryon are represented by the thick and thin lines, respectively.}
\label{fig:entro_05}
\end{figure}

From the internal energy and the entropy shown in the previous two
figures, one derives the free energy displayed in
Fig.~\ref{fig:free_05}. We observe that the hyperonic
contributions decrease the free energy at all temperatures and
densities. This effect is more pronounced at higher temperatures,
for which the negative entropic term of the free energy dominates.
However, for a given temperature the shift in the free energy per
particle is almost independent of the density.

\begin{figure}[htp] \centering
\includegraphics[width=9cm]{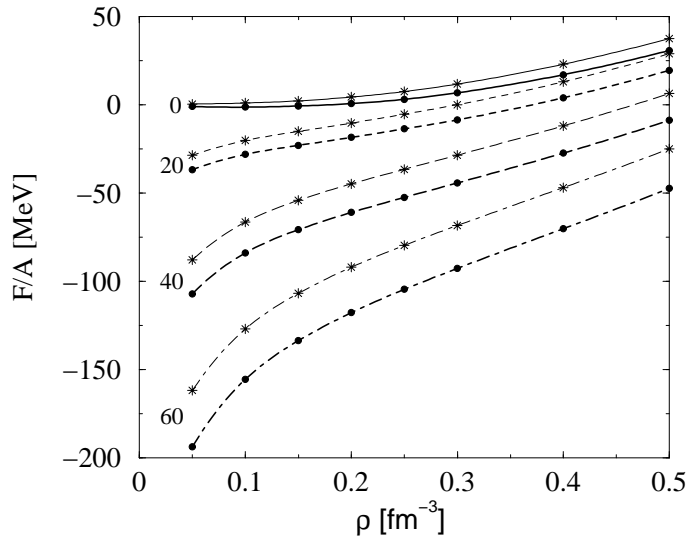}
\caption{Free energy per nucleon as a function of density of baryonic
matter composed by 80\% neutrons, 10\% protons and 10\% $\Sigma^-$
hyperons, for various temperatures:  0, 20, 40 and 60 MeV. The total 
(solid circles) and nucleonic-only (stars) contributions to the free energy 
per baryon are represented by the thick and thin lines, respectively.}
\label{fig:free_05}
\end{figure}


\section{Conclusions}
\label{sec:conclusions}

In the present work we have studied hadronic matter at finite
temperature and density within the NTBBG approach using the
Argonne V18 potential plus a three body force for the
nucleon-nucleon interaction and the Nijmegen NSC97e potential for
the hyperon-nucleon (YN) and hyperon-hyperon (YY) ones. To our
knowledge, this is the first time that a fully microscopic
calculation of  matter with hyperons is performed at finite
temperature.

In this approximation, the single-particle occupation
probabilities are replaced by thermal Fermi distributions. This
affects the Pauli blocking factor on the intermediate states
appearing in the equation that defines the in-medium effective
interaction. With increasing temperature, the Pauli blocking
looses strength and the interaction, which shows in general a
moderate temperature dependence, tends to acquire the behavior
observed for the corresponding free T-matrix.

We have obtained the single-particle potentials of all the baryons
for various temperatures in two different cases: symmetric nuclear
matter at normal density and a system composed by $80 \%$
neutrons, $10 \%$ protons and 10\% $\Sigma^-$ hyperons at a
density of $\rho=0.5$ fm$^{-3}$, which is representative of a
beta-stable composition of a neutron star. Temperature effects on
the single-particle potentials are at most of the order of $25 \%$
when the temperature increases from 0 to 60 MeV in all the
baryons, being more relevant for the imaginary parts, which
increase in size due to the gain in phase space. With increasing
temperature, the size of the real part of the potential at zero
momentum decreases (independently of whether it is attractive or
repulsive) due to the contribution of higher momentum components
for which the effective interaction is weaker.

We have also studied the dependence with density and temperature
of the bulk properties of hadronic matter, namely the energy,
entropy and free energy. 

In the case of symmetric nuclear matter
we have checked that our results are very similar to the previous
available calculations \cite{ba99}. Although the kinetic terms
play an increasing role with increasing temperatures, the final
value of the thermodynamical quantities is, in the range of
densities and temperatures explored in this work, substantially
affected by the interacting terms. We have compared the values of
the chemical potential obtained from the density normalization and
from the derivative of the free energy to have an estimate of the
rearrangement terms neglected in the NTBBG approach. It is found
that the rearrangement effects increase with density and decrease
with temperature. We note, however, that the 
the bulk thermodynamical quantities are  
little affected by these missing rearrangement components in the
single-particle spectrum \cite{ba99}. 

We have also analyzed the relevance of
hyperonic contributions to the bulk properties of matter by fixing
the composition to 80\% neutrons, 10\% protons and 10\% $\Sigma^-$
and varying the density and temperature. Due to the small fraction
of hyperons considered, the $\Sigma^-$ start behaving classically
for low densities. Thus the kinetic contribution to the energy
easily overcomes the $YN$ and $YY$ potential energy contribution.
However, the latter cannot be neglected since it represents about
half of the hyperonic kinetic contribution at the largest
temperature of $T=60$ MeV explored in this work. In addition, we
see that the entropy of the hadronic system is very sensitive to 
the presence of the $\Sigma$'s, despite their relatively low
fraction. Although the deviations with respect to the nucleonic
sector are more important for higher temperatures, they do not
depend on density. We also find that the modest presence of
hyperons leads to substantial reduction of the free energy,
obviously associated to the increase of the entropy.

As we have already pointed out, to our knowledge, this is the first
microscopic calculation of the finite temperature EoS of dense baryonic matter 
including degrees of freedom other than nucleons. It represents a first step 
in  the microscopic study of hadronic matter at finite temperature, which is 
of great importance, on one hand, for a proper understanding of the 
supernova and proto-neutron star physics and, on the other, for the analysis 
of heavy ion collisions data.


\section*{Acknowledgements}

We acknowledge fruitful discussions with Marcello Baldo.
This work is partly supported by DGICYT contract BFM2002-01868 and 
the Generalitat de Catalunya contract SGR2001-64.
This research is part of the EU Integrated Infrastructure Initiative
Hadron Physics Project under contract number RII3-CT-2004-506078. 
Arnau Rios acknowledges the support from DURSI and the European Social Funds.


\end{document}